\documentclass[aps,prl,twocolumn,superscriptaddress,amsmath,amssymb]{revtex4-1}

\usepackage{graphicx}% Include figure files
\usepackage{dcolumn}% Align table columns on decimal point
\usepackage{bm}% bold math
\usepackage[utf8]{inputenc} % allow utf-8 input
\usepackage[T1]{fontenc}    % use 8-bit T1 fonts
\usepackage{hyperref}       % hyperlinks
\usepackage{url}            % simple URL typesetting
\usepackage{booktabs}       % professional-quality tables
\usepackage{amsfonts}       % blackboard math symbols
\usepackage{nicefrac}       % compact symbols for 1/2, etc.
\usepackage{microtype}      % microtypography
\usepackage{natbib}
\usepackage{makecell}
\usepackage{comment}
\usepackage{xcolor}

\newcommand{\micron}{\hbox{\textmu}\text{m}}

\begin{document}

\title{Planar Bragg microcavities with monolayer WS$_2$ for strong exciton--photon coupling}

\author{Alexey O. Mikhin}
\affiliation{School of Physics and Engineering, ITMO University, St. Petersburg 197101, Russia}
\author{Albert A. Seredin}
\affiliation{School of Physics and Engineering, ITMO University, St. Petersburg 197101, Russia}
\author{Roman S. Savelev}
\affiliation{School of Physics and Engineering, ITMO University, St. Petersburg 197101, Russia}
\author{Dmitry N. Krizhanovskii}
\affiliation
{Department of Physics and Astronomy, University of Sheffield, Sheffield S3 7RH, UK}
\author{Anton K. Samusev}
\affiliation{ 
Experimentelle Physik 2, Technische Universität Dortmund, 44227 Dortmund, Germany}
\author{Vasily Kravtsov}
\email{vasily.kravtsov@metalab.ifmo.ru}
\affiliation{School of Physics and Engineering, ITMO University, St. Petersburg 197101, Russia}

%\date{\today}

\begin{abstract}
We propose and numerically investigate a novel compact planar microcavity design based on a high-index dielectric slab waveguide with embedded monolayer semiconductor.
In comparison to more traditional vertical Bragg microcavities, our design relies on the transmission of guided optical modes and achieves strong exciton--photon coupling in a chip-compatible and compact geometry with sub-100~nm thickness.
We show that Rabi splitting values of more than 70~meV can be obtained in planar microcavities with the total length below 5~\micron. 
Further, we reveal the dependence of Rabi splitting on the dimensions of the structure and explain it with a simple theoretical model.
Our results contribute towards the development of novel compact 2D semiconductor-based components for integrated photonic circuits.
\end{abstract}


\maketitle

%%%%%%%%%%%%%%%%%%%%%%%%%%%%%%%%%%%%%%%%%%%%%%%%%%%%%%%%%%%%%%%%%%%%%%%%%%%%%
\begin{figure*}[htp]
    \centering
    \includegraphics[width=\linewidth]{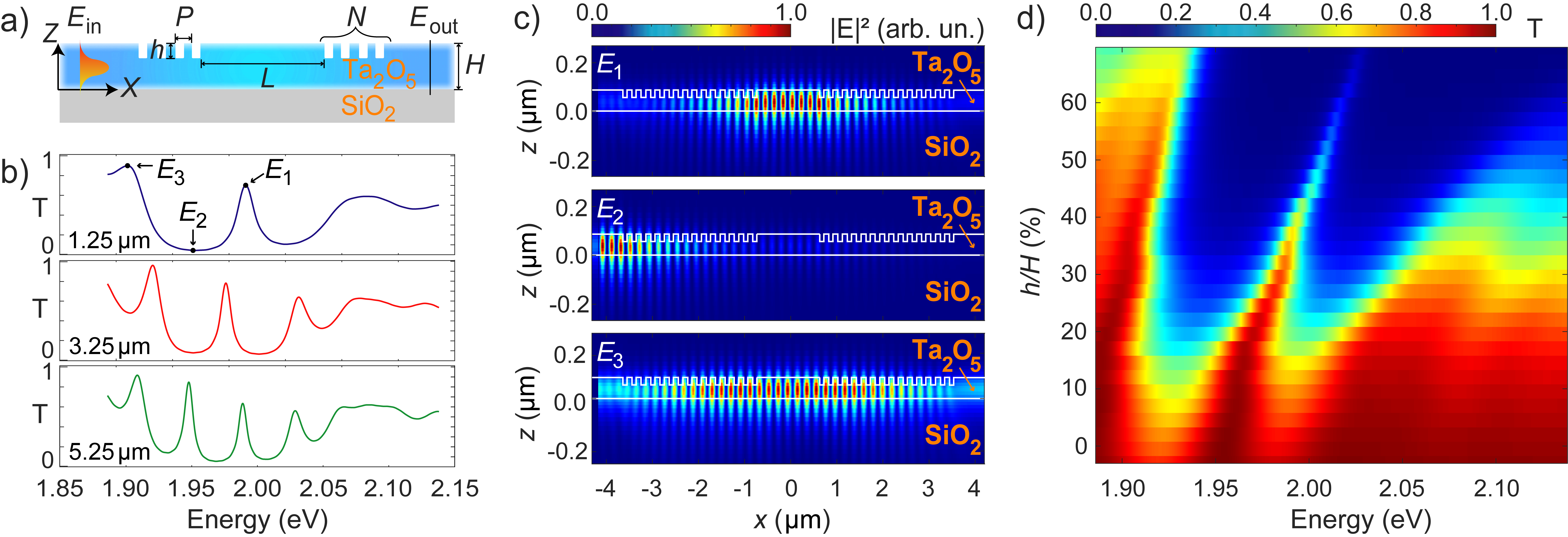}
    \caption{a) Schematic view of the studied planar Bragg microcavity. b) Transmittance spectra through the microcavity with number of DBR periods N~=~15 and different resonator length $L$: $1.25$~\micron~(top), $3.25$~\micron~(middle), and $5.25$~\micron~(bottom). c) Spatial distribution of in-plane magnitude of the electric field in a planar microcavity with $L=1.25$~\micron~and N~=~15 for 3 different photon energies: $\mathrm{E_1}=1.98$~eV (top), $\mathrm{E_2}=1.95$~eV (middle), and $\mathrm{E_3}=1.90$~eV (bottom) as indicated in panel (b). d) Transmittance spectra calculated for planar microcavities with various etching depths $h$ while maintaining a fixed thickness $H$ of the waveguide layer.}
    \label{fig:FIG_1}
\end{figure*}
%%%%%%%%%%%%%%%%%%%%%%%%%%%%%%%%%%%%%%%%%%%%%%%%%%%%%%%%%%%%%%%%%%%%%%%%%%%%%

%\section{\label{sec:level1}Introduction}
\noindent
Integrated nanophotonics holds great potential for applications in areas such as information processing, sensing, and energy conversion. Developing nonlinear and tunable nanophotonic structures is especially important as they allow on-chip manipulation and control of light at the nanoscale~\cite{Elshaari2020May, Wu2021Apr, Koo2024}, leading to enhanced performance and functionality of optical and optoelectronic systems. 
However, even materials exhibiting highly nonlinear and tunable optical properties~\cite{Gong2021Jan, Meng2023Aug} are limited in terms of nonlinear conversion efficiencies in on-chip structures due to small interaction volumes.

One promising approach to achieving enhanced nonlinear response in nanophotonic devices is the use of strong light--matter coupling regime~\cite{Khitrova1999Oct}, for example in semiconductor microcavities supporting exciton-polaritons~\cite{Skolnick1998Jul, Deng2002Oct}.
These hybrid quasi-particles exhibit unique optical properties, including strong light--matter interactions, ultrafast response times~\cite{Masharin2024Jan, Virgili2011Jun}, and the ability to undergo Bose-Einstein condensation~\cite{Kasprzak2006Sep, Plumhof2014Mar}.
Leveraging exciton-polaritons in nanophotonic devices opens up new possibilities to control and manipulate light at the nanoscale including all-optical switching~\cite{Feng2021Nov}, signal processing~\cite{Amo2010Jun}, and quantum information processing~\cite{Kavokin2022Jul}.

One of the recently emerged platforms that holds promise for achieving strong light--matter coupling and polaritonic effects is the two-dimensional transition metal dichalcogenides (TMDs) owing to their unique geometry and optical properties~\cite{Schneider2018Jul}.
In the monolayer (ML) form, TMDs are direct bandgap semiconductors hosting excitons with high oscillator strength and large binding energy~\cite{Chernikov2014Aug, Hsu2019Jul}.
Strong coupling of excitons in ML TMDs with photons can be achieved upon their integration into vertical microcavities formed by pairs of distributed Bragg reflectors (DBRs)~\cite{Liu2015Jan, Dufferwiel2015Oct}.
However, since configurations based on vertical DBR cavities are bulky and not well suited for integration on chips, there has been a lot of interest recently in studying more compact planar systems such as thin dielectric waveguides~\cite{Benimetskiy2023Aug, Kondratyev2023Sep}, photonic crystals~\cite{Zhang2018Feb, Khestanova2024Jun}, metasurfaces~\cite{Chen2020Jul, Li2021Jul}, and structures supporting Bloch surface waves~\cite{Barachati2018Oct} or bound states in the continuum~\cite{Kravtsov2020Apr}.
While such structures are significantly thinner than vertical DBR cavities, their in-plane dimensions normally range from 10~\micron~to 100~\micron, which calls for the development of new designs that can further reduce the footprint of polaritonic devices.
Typical quasi-zero-dimensional nanocavities explored in this context~\cite{li2017room, liu2017enhanced, Qian2022Jun, fang2022quantization, rosser2021optimal}, however, are characterized with increased fabrication complexity as well as intrinsically complex optical field distribution and lack of variable parameters for measuring polaritonic dispersion curves.
Consequently, this prevented demonstration of sizeable Rabi splitting values in such systems so far~\cite{rosser2022dispersive}.

Here, we propose and numerically investigate a novel design for planar micro-resonators based on a patterned subwavelength-thick dielectric slab waveguide coupled with monolayer TMD.
Modelling the transmission spectra for the waveguide mode incident on the micro-resonator, we reveal an anti-crossing behavior between optical and exciton resonances and show that the strong coupling regime with high Rabi splitting values exceeding 70~meV can be achieved for a wide range of system parameters.
Further, we find that both Rabi splitting $\Omega_\mathrm{R}$ and polariton linewidth $\Gamma$ decrease with increasing number of periods in the Bragg mirrors and suggest that the ratio $\Omega_\mathrm{R}/\Gamma$ can be used for optimization of the structure's parameters. 
Our results provide important insights for designing compact on-chip polaritonic elements based on the combined use of planar Bragg mirrors and 2D semiconductors. 
The increased interaction strength alongside the small mode volume in such structures can help boost the associated nonlinear optical response for developing next-generation active nanophotonic devices.

%\section{\label{sec:level2}Results}
To optimize parameters of the optical field for the strong light--matter coupling regime, we first consider a planar microcavity without a TMD monolayer.
A schematic of the microcavity is shown in Fig.~\ref{fig:FIG_1}a. 
The structure consist of a tantalum pentoxide (Ta$_2$O$_5$) slab waveguide of thickness $H=$~90~nm on top of a 1~\micron~thick SiO$_2$ substrate.
Due to the high refractive index of the Ta$_2$O$_5$ layer $n_\mathrm{T}=$~2.01 as compared to that of substrate $n_\mathrm{S}=$~1.46, the structure acts as a high-index dielectric waveguide supporting non-radiating photonic modes.
To create a microcavity for the waveguide modes, two Bragg mirrors are formed in the Ta$_2$O$_5$ layer at a distance $L$ away from each other via periodic rectangular modulation of the layer thickness with depth $h$ and number of periods $N$.
Experimentally, such patterning can be realized using reactive ion etching through a nanostructured metal mask.
The fill factor and pitch are chosen as $\eta = 0.5$ and $P=202$~nm, respectively, in order to achieve overlap between the DBR photonic bandgap and exciton resonance in monolayer WS$_2$.

We study the optical response of the planar microcavity by numerical simulation of the transmittance spectra for propagating waveguide modes using Lumerical FDTD software package.
As the incident wave, we use a TE mode propagating in the waveguide from the left as illustrated in Fig.~\ref{fig:FIG_1}a with $E_\mathrm{in}$.
After propagation through the microcavity, the transmitted wave is evaluated at the unpatterned region of the waveguide 0.5~\micron~to the right of the output Bragg mirror, which is indicated in the figure as $E_\mathrm{out}$. 
Transmittance spectra for microcavities with $N=15$ periods in each DBR and selected values of microcavity length $L=1.25$~\micron, $3.25$~\micron, and $5.25$~\micron~are shown in Fig.~\ref{fig:FIG_1}b in the top, middle, and bottom panels, respectively.
The transmittance spectra exhibit narrow resonance peaks within a broader photonic stop-band, with the exact spectral shape sensitively depending on the microcavity parameters. 
The width of the resonance peaks is determined by the number of DBR  periods $N$, where larger $N$ results in narrower peaks and correspondingly increased Q-factor.
At the length L = 1.25~\micron, an increase in N from 5 to 15 results in a reduction of the photonic mode linewidth from 44~meV to 8~meV, subsequently enhancing the Q-factor of the microcavity from 45 to 240.
Increasing microcavity length $L$ leads to a multi-mode structure within the stop-band and a corresponding decrease in finesse~\cite{Yablonovitch1987May}. 

Fig.~\ref{fig:FIG_1}c shows the simulated electric field distribution expressed as the square of the field magnitude $|E_{||}|^2$ along the $x$- and $y$-directions for a microcavity with $L=1.25$~\micron~and $N=15$ at three different photon energies indicated in the top panel of Fig.~\ref{fig:FIG_1}b.
For the photon energy corresponding to the peak in the transmittance spectrum ($\mathrm{E_1}=1.98$~eV, top panel), the field is predominantly localized inside the cavity between the DBRs.
For the photon energy in the stop-band ($\mathrm{E_2}=1.95$~eV, middle panel), the incident waveguide mode experiences reflection and scattering off the DBRs.
Outside the stop-band ($\mathrm{E_3}=1.90$~eV, bottom panel), the incident wave is efficiently transmitted through the microcavity due to excitation of the resonances that are predominantly localized in the Bragg mirrors. %without additional enhancement between the DBRs.

%%%%%%%%%%%%%%%%%%%%%%%%%%%%%%%%%%%%%%%%%%%%%%%%%%%%%%%%%%%%%%%%%%%%%%%%%%%%%
\begin{figure*}[htp]
    \centering
    \includegraphics[width=\textwidth]{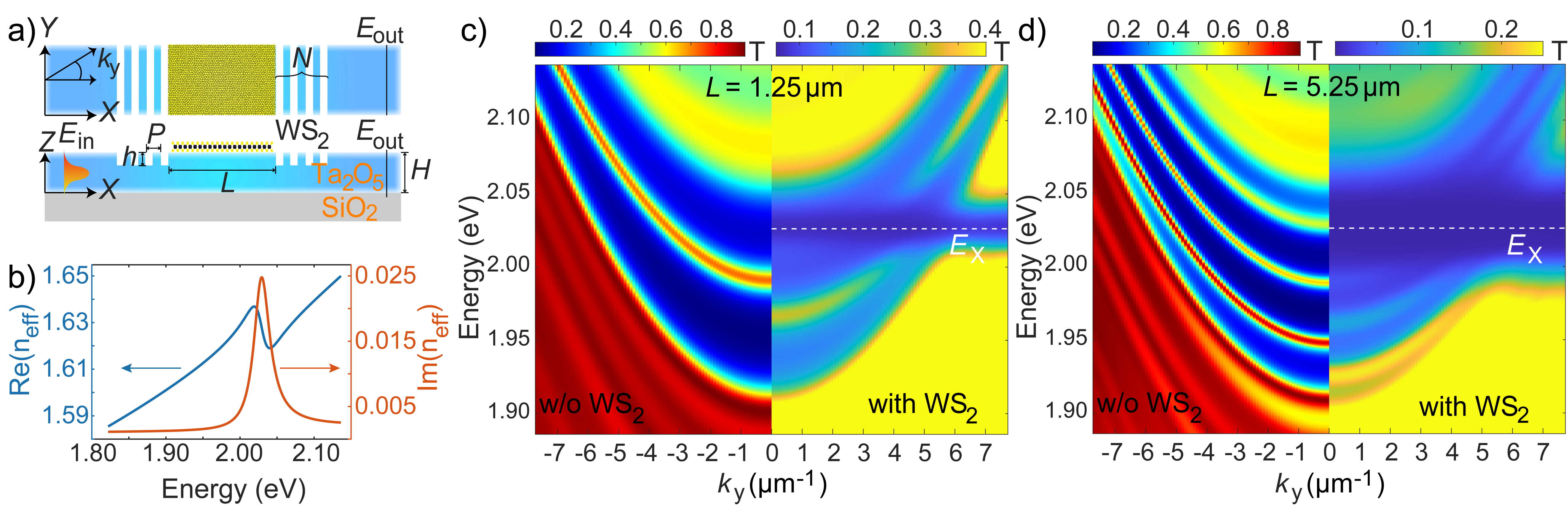}
    \caption{a) Schematic view of a planar Bragg microcavity with a WS$_2$ monolayer integrated between DBRs. b) Real and imaginary parts of the effective refractive index for the hybrid waveguide structure. c) Transmittance spectra for a single-mode microcavity of length $L=1.25$~\micron~without (left side) and with (right side) WS$_2$ monolayer for different wavevector components $k_y$. d) Corresponding transmittance spectra for a multi-mode microcavity of length $L=5.25$~\micron .}
    \label{fig:FIG_2}
\end{figure*}
%%%%%%%%%%%%%%%%%%%%%%%%%%%%%%%%%%%%%%%%%%%%%%%%%%%%%%%%%%%%%%%%%%%%%%%%%%%%%

In contrast to traditional vertical DBR microcavities, the transmission in our planar structure, as well as the corresponding electric field magnitude and Q-factor, can be controlled via modulation depth $h$.   
To optimize this parameter, we calculate transmittance spectra for different values of $h$ at a fixed waveguide thickness $H=90$~nm. 
The results are plotted in Fig.~\ref{fig:FIG_1}d, where the modulation depth is shown on the vertical axis as a fraction of waveguide thickness $H$.
With increasing modulation depth, the photonic stop-band becomes more pronounced, while the Q-factor of the resonance increases and the corresponding transmission coefficient decreases.
For further analysis, we choose $h/H = 1/3$ as this value maximizes the ratio between the transmittance and linewidth of the resonance peak, which then results in optimized light--matter coupling.

Next, we consider a hybrid planar microcavity with an embedded monolayer semiconductor WS$_2$ placed between the DBRs without overlap with the mirrors, as schematically illustrated in Fig.~\ref{fig:FIG_2}a. 
To perform simulations, the monolayer is incorporated into the computational domain as a three-dimensional sample object with a thickness of $0.62$~nm and dielectric function derived from~\cite{Li2014Nov}. 
The real part of the resulting effective refractive index of the structure, which consists of WS$_2$ on the Ta$_2$O$_5$ and SiO$_2$ substrate, is presented in Fig.~\ref{fig:FIG_2}b by the blue line, while the red line corresponds to the imaginary part.

To obtain the energy--wavevector dispersion of the studied planar microcavity, we consider waveguide modes with different $k_y$ components of the wavevector and simulate transmittance spectra using Bloch boundary conditions along the $y$-direction.
The resulting spectra for hybrid planar microcavities with resonator lengths of $L=1.25$~\micron~and $5.25$~\micron~are plotted in the right panels of Fig.~\ref{fig:FIG_2}c and Fig.~\ref{fig:FIG_2}d, respectively.
For comparison, we plot corresponding transmittance spectra for bare microcavities without monolayer WS$_2$ in the left panels of Fig.~\ref{fig:FIG_2}c,d.

%%%%%%%%%%%%%%%%%%%%%%%%%%%%%%%%%%%%%%%%%%%%%%%%%%%%%%%%%%%%%%%%%%%%%%%%%%%%%
\begin{figure*}[htp]
    \centering 
    \includegraphics[width=1\textwidth]{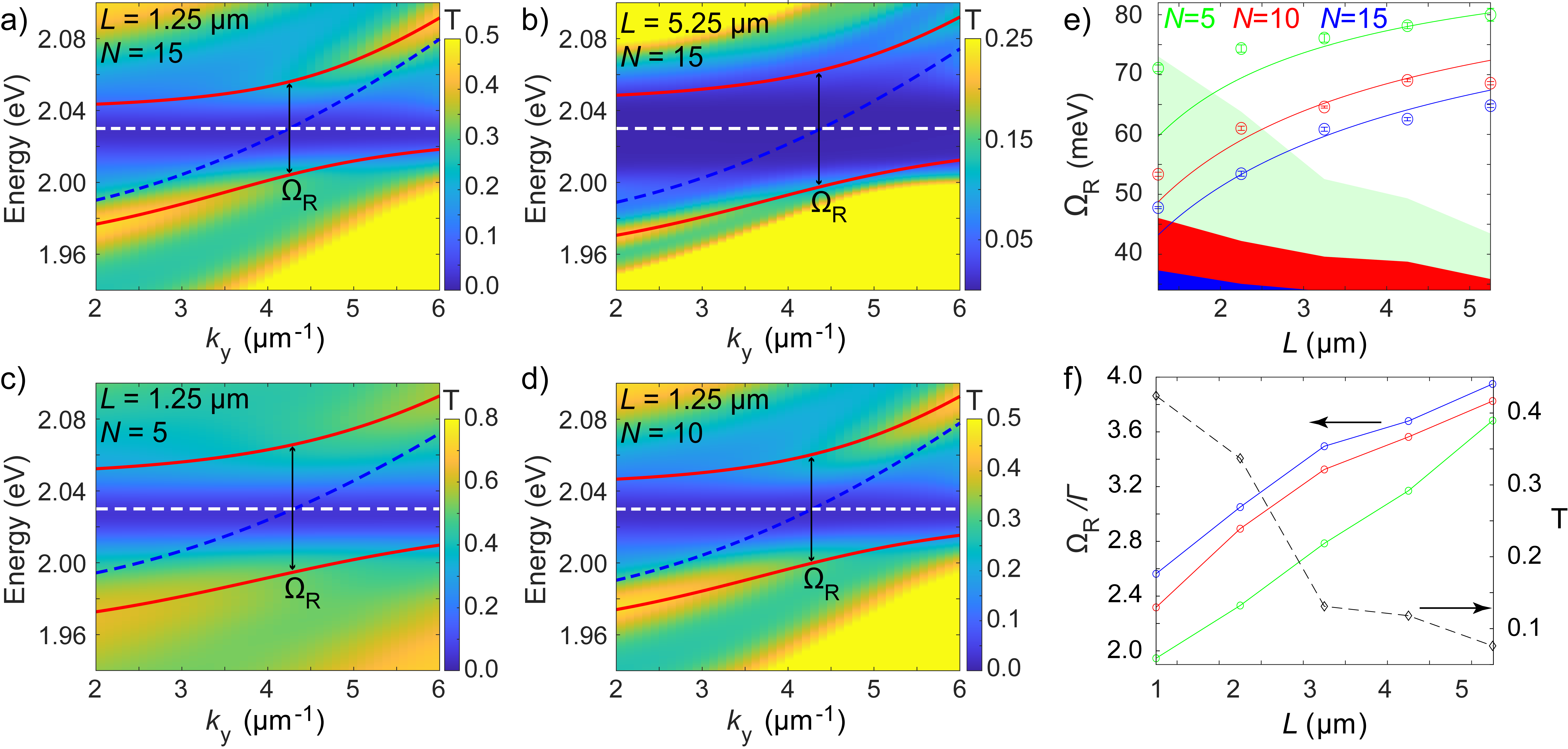}
    \caption{a-d) Approximation of numerical transmittance spectra by coupled oscillator model for structures with different resonator length $1.25$~\micron~(a) and $5.25$~\micron~(b), and for structures with different number of DBR periods 5 (c) and 10 (d). Cavity, exciton, and polariton dispersions are indicated by blue dashed, white dashed, and red solid lines. e) Extracted Rabi splitting values for varying length of the resonator: simulated (points) and analytical (solid curves). Colors indicate the number of DBR periods: 5 (green), 10 (red), and 15 (blue). Colored dashed areas indicate the onset of the weak coupling regime. f) Ratio of Rabi splitting to polariton linewidth (left axis) and transmittance vs. resonator length for different number of DBR periods (right axis).}
    \label{fig:FIG_3}
\end{figure*}
%%%%%%%%%%%%%%%%%%%%%%%%%%%%%%%%%%%%%%%%%%%%%%%%%%%%%%%%%%%%%%%%%%%%%%%%%%%%%

Integration of a WS$_2$ ML into the microcavity results in dispersion splitting around the exciton resonance ($\mathrm{E_{X}}=2.03$~eV, white dashed line) and formation of lower and upper polariton (LP and UP) branches as seen in the right panels of Fig.~\ref{fig:FIG_2}c,d. 
In order to numerically describe the interaction between excitons in monolayer WS$_2$ and photons in the planar microcavity, we fit the simulated dispersion data by the coupled oscillator model~\cite{Schneider2018Jul}: %where the exciton--photon coupling strength $g$ is a fitting parameter:
\begin{equation} \label{eq1}
    \begin{split}
        \mathrm{E_{UP(LP)}}(k_y) = \frac{\mathrm{E_{X}}+\mathrm{E_{C}}(k_y)}{2}-i\frac{\gamma_\mathrm{X}+\gamma_\mathrm{C}}{2}\\ \pm \frac{1}{2}\sqrt{[\mathrm{E_{C}}(k_y)-\mathrm{E_{X}}-i(\gamma_\mathrm{C}-\gamma_\mathrm{X})]^2+4g^2},
    \end{split}
\end{equation}
where $\mathrm{E_{C}}$ and $\gamma_\mathrm{C}$ are the energy and decay rate of the photonic mode, $\gamma_\mathrm{X}$ is the exciton decay rate, and $g$ is the exciton--photon coupling strength employed as the main fitting parameter.
Furthermore, integration of WS$_2$ ML with the microcavity modifies the dielectric environment, which we account for by introducing two additional fitting parameters describing a shift and slope in the photonic dispersion.
To perform fits, we obtain the polaritonic $\mathrm{E_{UP(LP)}}$ and photonic $\mathrm{E_C}$ dispersions at different $k_y$-vectors by fitting the corresponding resonance peaks in transmittance spectra with Lorentzian line shapes. 
The polaritonic $\Gamma$ and photonic $\gamma_\mathrm{C}$ decay rates are obtained as corresponding Lorentzian linewidths, while excitonic decay rate $\gamma_\mathrm{X}=29$~meV is estimated from the width of the resonance in dielectric function shown in Fig.~\ref{fig:FIG_2}b.

Using the extracted from fits exciton--photon coupling strength $g$, we calculate the Rabi splitting energy as follows:
\begin{equation}
\Omega_\mathrm{R}= \sqrt{4g^2-(\gamma_\mathrm{C}-\gamma_\mathrm{X})^2}.
\end{equation}
Selected results of fitting are plotted in Fig.~\ref{fig:FIG_3}a-d, where the photonic and polaritonic dispersions are shown with blue dashed and red solid curves, respectively, and the exciton resonance is shown with white dashed lines.
The corresponding Rabi splitting energies $\Omega_\mathrm{R}$ are indicated with black double-headed arrows.
Panels (a) and (b) in Fig.~\ref{fig:FIG_3} correspond to the same DBR geometry ($N=15$) but different resonator length $L=1.25$~\micron~and $5.25$~\micron.
For larger $L$, the dispersion shows a multi-mode structure and an increased Rabi splitting.
We note that, when fitting data for multi-mode resonators, we select the mode that exhibits a resonance frequency close to that of the single-mode case.
Panels (c) and (d) in Fig.~\ref{fig:FIG_3} correspond to the same resonator length $L=1.25$~\micron~but different DBR geometry with $N=5$ and $N=10$, respectively.
These dispersions exhibit a single-mode character and reveal a counter-intuitive decrease of both Rabi splitting and coupling strength $g$ for increasing $N$.

Various values of Rabi splitting obtained from fits of the simulated transmittance spectra for hybrid planar microcavities with different parameters are summarized in Fig.~\ref{fig:FIG_3}e.
Symbols of different colors correspond to microcavities with different number of periods in their DBRs (green for 5 periods, red for 10 periods, and blue for 15 periods), and the horizontal axis corresponds to the varying microcavity length $L$.
The presented data demonstrate two distinct trends.
First, the values of Rabi splitting increase for increasing resonator length, with saturation at large values of $L$.
Second, for resonators of the same length, the values of Rabi splitting decrease for increasing number of DBR periods.
In the following, we discuss these two trends.

%\section{\label{sec:level3}Discussion}
Similar to the case of traditional vertical Bragg cavities~\cite{McGhee2021Oct}, the strength of light--matter coupling in our structure is defined by two factors: (i) degree of the photonic mode confinement in the microcavity and (ii) overlap between the photonic mode and excitons in the WS$_2$ monolayer.
The photonic mode confinement can be described by its mode volume, which in our case is represented by the effective length of the microcavity in the $x$-direction, that is, the extent over which the electromagnetic field is effectively confined within the cavity.
To evaluate the overlap between the photonic mode and excitons, the effective length $L_\mathrm{eff}$ can be compared to the length of the WS$_2$ monolayer, which in our case is equal to the resonator length $L$.

When the resonator length increases without change in the geometry of Bragg mirrors, both the WS$_2$ monolayer length and effective cavity length increase, with their ratio $L/L_\mathrm{eff}$ increasing and approaching unity for large $L$.
This explains the observed dependence of Rabi splitting on resonator length, which exhibits an initial increase followed by saturation as seen in Fig.~\ref{fig:FIG_3}e.
When the number of periods in Bragg mirrors increases without change in resonator length $L$, it results in increased effective cavity length and decreased ratio $L/L_\mathrm{eff}$.
This leads to a reduced photonic mode localization and reduced overlap between the photonic mode and WS$_2$ excitons, which results in the observed decrease of Rabi splitting for increasing number of periods in Bragg mirrors $N$.

To provide a simple model for the observed trends of Rabi splitting $\Omega_\mathrm{R}$, we calculate exciton--photon coupling strength by integrating the electric field of the cavity eigenmode over the volume of the TMD layer~\cite{gerace2007quantum}. 
To obtain a simple analytical result, we approximate the photonic eigenmode field as a combination of properly normalized standing wave inside the cavity and the exponentially decaying waves in the mirrors. 
Further, simplification can be made by assuming that the decay length of the wave in the mirrors is much larger than the mirror period. Such condition is well satisfied in the considered structures. 
As a result, we obtain 
\begin{equation}
\Omega_\mathrm{R} \propto \sqrt{\frac{1}{1+\frac{1-\mathrm{exp} \left( -2 \alpha N P \right)}{\alpha L}}},
\end{equation}
where $P$ is the period of the mirrors, and $\alpha$ is the field decay constant in the mirrors.
We calculate the analytical values of Rabi splitting as a function of the resonator length at different numbers of periods in Bragg mirrors and plot them in Fig.~\ref{fig:FIG_3}e as solid curves.
As can be observed, the analytical results provide a qualitative description of the behavior observed in the dependence extracted from the simulated polaritonic dispersion.

For potential future experimental investigation, it is important to determine whether the studied structures support the strong light--matter coupling regime.
To this end, we compare the obtained coupling strength $g$ with excitonic and photonic linewidth.
The strong coupling regime occurs for $g > \sqrt{(\gamma_\mathrm{X}^2+\gamma_\mathrm{C}^2)/2}$.
To check this condition, we calculate the parameter $\gamma_\mathrm{eff} = \sqrt{(\gamma_\mathrm{X}^2+\gamma_\mathrm{C}^2)/2}$ and plot it in Fig.~\ref{fig:FIG_3}e as shaded-area curves with colors corresponding to different number of periods in Bragg mirrors: $N = 5$ (green), $N = 10$ (red), and $N = 15$ (blue).
As seen in the figure, the strong coupling regime is achieved for almost all studied parameters, except for the structure with $N = 5$ and $L = 1.25$~\micron.

Since both the Rabi splitting and polariton linewidth depend on the resonator length and Bragg mirror geometry, we propose their ratio $\Omega_\mathrm{R} / \Gamma$ as the parameter for optimizing design of experimental polaritonic structures.
In our model, $\Gamma$ is evaluated on the polariton resonance as a half-sum of the uncoupled exciton linewidth and the photonic mode linewidth: $\Gamma=(\gamma_\mathrm{X}+\gamma_\mathrm{C})/2$~\cite{Dufferwiel2015Oct}.
The resulting ratio $\Omega_\mathrm{R} / \Gamma$ is plotted in Fig.~\ref{fig:FIG_3}f (colored symbols, left scale) and exhibits increase with both $L$ and $N$.
We note, however, that for experimental realizations of strongly-coupled planar resonators, the overall transmittance through the structure is an important consideration.
The calculated transmittance is plotted in Fig.~\ref{fig:FIG_3}f with black symbols (right scale) and exhibits a decrease for increasing $L$.
Therefore, a trade-off between the increasing ratio $\Omega_\mathrm{R} / \Gamma$ and decreasing overall transmittance should be considered when designing experimental polaritonic structures.
Our calculations show that optimal parameters lie in the range $1.5-5.0$~\micron~for the resonator length $L$ and in the range of $5-15$ for the number of periods $N$ in Bragg mirrors.

%\section{\label{sec:level3}Conclusions}
In summary, we have proposed and numerically investigated a novel type of compact photonic resonator supporting the strong light--matter coupling regime.
The resonator consists of a high-index dielectric slab waveguide with patterned Bragg mirrors interfaced with monolayer WS$_2$ and can be made as small as sub-100~nm in thickness and sub-5~\micron~in length while still exhibiting formation of distinct exciton--polaritons with Rabi splitting in excess of 70~meV.
We have found that Rabi splitting for such hybrid structure increases with increasing resonator length and decreases with increasing number of periods in Bragg mirrors.
To explain both trends, we have provided a simple model that accounts for the interplay between the resonator length $L$ and effective length $L_\mathrm{eff}$, which quantifies the optical mode localization.
Our results provide important insights for future development of compact and chip-compatible resonant photonic structures based on 2D semiconductors operating under strong light--matter coupling.
Characterized by sizeable polariton--polariton interaction and the possibilities of electrostatic control, such structures could be particularly promising candidates as logic gates in integrated photonic circuits.

\begin{acknowledgments}
This work was supported by Priority 2030 Federal Academic Leadership Program.
\end{acknowledgments}

%\newpage
%\section{\label{sec:level4}References}
\bibliographystyle{naturemag}
\bibliography{planarbragg} 

\end{document}